\documentclass[aps,showpacs,twocolumn,amsmath,amssymb,superscriptaddress]{revtex4-1}
\usepackage[colorlinks=true, citecolor=blue, urlcolor=blue ]{hyperref}
\usepackage{graphicx}
\usepackage{subfigure}
\usepackage{grffile}
\usepackage{bm}
\usepackage[version=3]{mhchem}
\usepackage{color}
\usepackage{hyperref}
\usepackage[normalem]{ulem}

\newcommand{\abs}[1]{\ensuremath{\left| #1 \right|}}
\newcommand{\ket}[1]{{\vert #1\rangle}}

\newcommand{\braket}[2]{\langle#1\vert#2\rangle}
\newcommand{\eval}[3]{\langle#1\vert#2\vert#3\rangle}

\newcommand{\1}{\mbox{\bf 1}}
\newcommand{\ud}{\mathrm{d}}
\newcommand{\RE}{\text{Re}}

\begin{document}

\title{Analysis of time-resolved single-particle spectrum on the one-dimensional extended Hubbard model}

\author{Can Shao}
\email{shaocan2018@csrc.ac.cn}
\affiliation{Beijing Computational Science Research Center, Beijing 100084, China}

\author{Takami Tohyama}
\affiliation{Department of Applied Physics, Tokyo University of Science, Tokyo 125-8585, Japan}


\author{Hong-Gang Luo}
\affiliation{School of Physical Science and Technology $\&$ Key Laboratory for Magnetism and Magnetic Materials of the MoE, Lanzhou University, Lanzhou 730000, China}
\affiliation{Beijing Computational Science Research Center, Beijing 100084, China}

\author{Hantao Lu}
\email{luht@lzu.edu.cn}
\affiliation{School of Physical Science and Technology $\&$ Key Laboratory for Magnetism and Magnetic Materials of the MoE, Lanzhou University, Lanzhou 730000, China}

\date{\today}

\begin{abstract}
We investigate the short-time evolution of the half filled one-dimensional extended Hubbard model in the strong-coupling regime, driven by a transient laser pump. Combining twisted boundary conditions with the time-dependent Lanczos technique, we obtain snapshots of the single-particle spectral function with high momentum resolution. The analysis of the oscillations of the spectral function shows that its characteristic frequencies are consistent with the magnitudes of the optical gap. Furthermore, we examine the time-evolving spectral structure in the charge-density-wave phase in detail, and find that one of the bands in the single-particle spectrum originates from the photoinduced bond-order background.
\end{abstract}


\maketitle

\section{Introduction}
\label{sec1}

Investigations on nonequilibrium processes can provide new enlightening information on dynamical properties of strongly correlated systems. Well-known examples are the pump-probe optical measurements that can unravel to some extent the complex entanglement of various degrees of freedom on ultrafast time-scales. This is largely attributed to the development of time-resolved spectroscopy techniques, such as time- and angle-resolved photoemission spectroscopy (trARPES), and transient transmissivity and reflectivity measurements.

In addition, photo-irradiation techniques, including the application of either strong electric fields or transient laser pulses, can lead to the observation of rich dynamics, such as insulator-to-metal and even insulator-to-superconductor transitions~\cite{Iwai03, Oka03, Oka05, Okamoto07, Takahashi08, Eckstein10, Kaneko19}. In underdoped cuprates, light-induced and light-enhanced superconductivity has attracted much attention as well~\cite{Fausti11, Nicoletti14, Hu14}. More recently, the optical enhancement of bond-order wave has been investigated on the extended Hubbard model~\cite{Shao19, Shinjo19}. We note that by tuning the parameters of the pumping pulse, the emergent dimerization can be of purely electronic origin, i.e., it is not associated with electron-lattice couplings as observed in alkali-TCNQ compounds~\cite{McQueen09,Uemura12}.


The trARPES is becoming a powerful tool for examining ultrafast phenomena in strongly correlated systems, such as gap collapse and reformation after photoexcitation~\cite{Rohwer11, Petersen11, Perfetti06, Perfetti08}. However, theoretical and numerical investigations of trARPES are quite demanding and challenging due to the dual difficulty presented by quantum correlation and nonequilibrium effects. At present, there are mainly two ways to obtain time-resolved single-particle spectra that can be compared with trARPES data. One is to use nonequilibrium Green's functions, i.e., the Keldysh formalism, to address the nonequilibrium issue. The correlation effects are tackled by combining other methods, e.g., the equations of motion or the dynamical mean-field theory~\cite{Freericks2006, Freericks2009, Aoki14, Kemper15, Nosarzewski17, Kemper17}. An alternative and perhaps more straightforward way relies on unbiased numerical methods, for instance, the time-dependent exact diagonalization (ED). However due to the severe constraint on system size in ED, the (time-dependent) correlation functions in the momentum space can only be defined at discrete points if the periodic boundary condition (BC) is imposed~\cite{Kanamori09}. One way to overcome this discreteness is to introduce twisted BCs, which have already been widely used in equilibrium ED calculations, e.g., see Refs.~\onlinecite{Poilblanc91,Tsutsui96,Tohyama04}.



In this paper, we propose an ED method that enables us to obtain the time-dependent single-particle spectral function with high momentum resolution. We then employ it to investigate the nonequilibrium pump-probe properties of the one-dimensional extended Hubbard model (1D EHM) at half filling, where the ground-state phase diagram in the strong-coupling regime is composed of two phases: a spin-density wave (SDW) phase and a charge-density wave (CDW) phase. We find that in both phases, the oscillation frequency of the spectral weight in the time-dependent single-particle spectral function is consistent with the magnitude of the optical gap.
Further in the CDW phase, a photoinduced enhancement is observed in one of two electron-removal bands, accompanied with the appearance of the elusive bond-order-wave (BOW) order~\cite{Nakamura00,Tsuchiizu02,Sengupta02,Jeckelmann02,Sandvik04,Ejima07} in the photoexcited state~\cite{Shao19}. This phenomenon may serve as a characteristic indication of the existence of the hidden BOW order in the excited states of the system.

The rest of the paper is organized as follows. In Sec.~\ref{sec_model}, we introduce the model and numerical methods employed to calculate the time-dependent single-particle spectral function and the optical conductivity. After discussions on equilibrium features, including those near a SDW-CDW boundary, analyses on nonequilibrium dynamics driven by a pumping pulse is carried through in Sec.~\ref{sec_pump}. The conclusion is given in Sec.~\ref{sec_conclusion}.

\section{Model and Method}\label{sec_model}

The 1D EHM at half filling reads
\begin{eqnarray}
H&=&-t_h\sum_{i,\sigma}\left(c^{\dagger}_{i,\sigma} c_{i+1,\sigma}+\text{H.c.}\right)+U\sum_{i}\left(n_{i,\uparrow}-\frac{1}{2}\right)\nonumber\\
&&\times\left(n_{i,\downarrow}-\frac{1}{2}\right)
+V\sum_{i}\left(n_{i}-1\right)\left(n_{i+1}-1\right),
\label{H}
\end{eqnarray}
where $c^{\dagger}_{i,\sigma}$ ($c_{i,\sigma}$) is the creation (annihilation) operator of an electron at site $i$ with spin $\sigma=\uparrow,\downarrow$, and the number operator of electrons $n_i= n_{i,\uparrow}+ n_{i,\downarrow}$, $t_h$ is the hopping constant, and $U$ and $V$ are the on-site and nearest-neighbor Coulomb repulsion strengths, respectively. In the rest of the paper, we use units with $e=\hbar=c=1$, and the lattice spacing $a_0=1$. In these units, $t_h$ and ${t_h}^{-1}$ are set to be the unit of energy and time, respectively.

Throughout the paper, we restrict ourselves to zero temperature. We are especially interested in the single-particle spectral function $I(k,\omega)$ and its time-dependent version $I(k,\omega,t)$. Note that, for a 1D chain of length $L$, if the standard periodic boundary condition is used, the set of allowed momenta in the first Brillouin zone reads $\mathcal{K}=\{k_l=2\pi l/L, l=0, 1, \ldots, L-1\}$. To estimate the spectral value at other momenta, e.g., $k_l+\kappa$ with $\kappa$ differing from the given set, a twisted BC can be employed, which equivalently results in the following transformation on the Hamiltonian (\ref{H})~\cite{Poilblanc91, Tsutsui96, Tohyama04}:
\begin{eqnarray}
c^{\dagger}_{i,\sigma} c_{i+1,\sigma}+\text{H.c.} \rightarrow
e^{\mathrm{i}{\kappa}}c^{\dagger}_{i,\sigma} c_{i+1,\sigma}+\text{H.c.}.
\label{eq:twist}
\end{eqnarray}
Note that the electronic operators still satisfy the periodic BC, i.e., $c_{i+L,\sigma}=c_{i,\sigma}$ for instance.

We now move to the detailed expressions of the single-particle spectral function $I(k,\omega)$. First we separate $I(k,\omega)$ into two parts:
\begin{eqnarray}
I(k,\omega)=I_{+}(k,\omega)+I_{-}(k,\omega),
\label{A}
\end{eqnarray}
with
\begin{eqnarray}
&I_{+}(k,\omega)=\sum\limits_{m,\sigma}\abs{\eval{\Psi_m^{\kappa}}{c_{k_*,\sigma}^{\dag}}{\Psi_0^{\kappa}}}^{2} \delta(\omega-(E_m^{\kappa}-E_0^{\kappa})-\mu_{\kappa}) \nonumber \\
&=-\frac{1}{\pi}\text{Im}\left(\sum\limits_{\sigma}\eval{\Psi_0^{\kappa}}{c_{k_*,\sigma}\frac{1}{\omega-(H^{\kappa}-E^{\kappa}_0)-\mu_{\kappa}-\mathrm{i}\eta}c_{k_*,\sigma}^{\dag}}{\Psi_0^{\kappa}}\right)
\label{A+}
\end{eqnarray}
and
\begin{eqnarray}
&I_{-}(k,\omega)=\sum\limits_{m,\sigma}\abs{\eval{\Psi_m^{\kappa}}{c_{k_*,\sigma}}{\Psi_0^{\kappa}}}^{2} \delta(\omega+(E_m^{\kappa}-E_0^{\kappa})-\mu_{\kappa}) \nonumber \\
&=-\frac{1}{\pi}\text{Im}\left(\sum\limits_{\sigma}\eval{\Psi_0^{\kappa}}{c_{k_*,\sigma}^{\dag}\frac{1}{\omega+(H^{\kappa}-E^{\kappa}_0)-\mu_{\kappa}-\mathrm{i}\eta}c_{k_*,\sigma}}{\Psi_0^{\kappa}}\right).
\label{A-}
\end{eqnarray}
$I_{+}$ ($I_{-}$) is the so-called electron-addition (electron-removal) spectral function, where $c_{k_*,\sigma}^{\dag}$ ($c_{k_*,\sigma}$) is the Fourier transformation of $c_{i,\sigma}^{\dag}$ ($c_{i,\sigma}$) at $k_*\in\mathcal{K}$. The resulting momentum $k$ is given by $k_*$ with a displacement of $\kappa\notin\mathcal{K}$, i.e., $k=k_*+\kappa$. Regarding a given $\kappa$ and the corresponding Hamiltonian $H^{\kappa}$ whose hopping terms are modified in Eq.~(\ref{eq:twist}), $\Psi_0^{\kappa}$ and $\Psi_m^{\kappa}$ represent the ground state and an intermediate $m$-state, with energy $E^{\kappa}_0$  and $E^{\kappa}_m$, respectively. Note that we always have the chemical potential $\mu_{\kappa}=0$ since only the half filling case is considered throughout the paper. $\eta$ is the spectral broadening factor and set to be $0.2$ in the calculations.

The external electric field during photoirradiation can be included into the Hamiltonian via the Peierls substitution in the hopping terms:
\begin{equation}
c^{\dagger}_{i,\sigma}c_{i+1,\sigma}+\text{H.c.}\rightarrow
e^{\mathrm{i}A(t)}c^{\dagger}_{i,\sigma}c_{i+1,\sigma}+\text{H.c.},
\label{eq:Peierls}
\end{equation}
where $A(t)$ is the vector potential of the applied (spatially uniform) ultrafast pulse. In this paper we use the form
\begin{equation}
A(t)=A_0e^{-\left(t-t_0\right)^2/2t_d^2}\cos\left[\omega_0\left(t-t_0\right)\right]
\label{eq:vpotent}
\end{equation}
to imitate the ultrafast pulses, where the temporal envelope of $A(t)$ centered at $t_0$ is taken to be Gaussian. The parameter $t_d$ controls its width, and $\omega_0$ is the central frequency of the electromagnetic wave.

The Peierls substitution in Eq.~(\ref{eq:Peierls}) can be generalized to incorporate the twisted BC imposed in Eq.~(\ref{eq:twist}):
\begin{equation}
c^{\dagger}_{i,\sigma}c_{i+1,\sigma}+\text{H.c.}\rightarrow
e^{\mathrm{i}A(t)}e^{\mathrm{i} \kappa}c^{\dagger}_{i,\sigma}c_{i+1,\sigma}+\text{H.c.}.
\label{eq:A}
\end{equation}
The generalization enables us to simulate the evolution starting from the $\kappa$-dependent initial state $\ket{\Psi^{\kappa}(0)}$ under the influence of $A(t)$ [the resulting  state is simply denoted as $\ket{\Psi^{\kappa}(t)}$], by employing the standard time-dependent Lanczos method~\cite{Prelovsek}. Then, by the substitution of $\ket{\Psi_0^{\kappa}}$ in Eqs.~(\ref{A+}) and (\ref{A-}) with $\ket{\Psi^{\kappa}(t)}$, and $H^{\kappa}$ with $H^{\kappa}(t)$, the quantity $I(k,\omega,t)$, which measures the single-particle excitation with respect to nonequilibrium states, can be calculated. Note that $E_0^{\kappa}$ in Eqs.~(\ref{A+}) and (\ref{A-}) should also be replaced by $E^{\kappa}(t)=\eval{\Psi^{\kappa}(t)}{H}{\Psi^{\kappa}(t)}$ in time-dependent calculations.

In this paper, in addition to the single-particle spectrum $I(k,\omega,t)$, the time-resolved optical conductivity $\sigma(\omega,t)$ is also investigated. To obtain $\sigma(\omega,t)$, we adopt the method derived rigorously from linear-response theory, which is in fact equivalent to the pump-probe method with $\delta$-like probing pulse~\cite{Zala14, Shao16}. Here the time-resolved optical conductivity is defined as
\begin{equation}
\sigma(\omega,t)=\int_0^{+\infty}\sigma(t+s,t)e^{i(\omega+i\delta)s}\,\ud s.
\label{eq:sigmaomega}
\end{equation}
In our numerical simulations, the integration cutoff for $s$ is taken to be several hundred time units. The broadening factor $\delta$ is set to be $1/L$. The response function $\sigma(t',t)$ (with restriction $t'\ge t$) reads
\begin{equation}
\sigma(t',t)=\frac{1}{L}\left[\eval{\psi(t')}{\tau}{\psi(t')}+\int_t^{t'}\chi(t',t'')\,\ud t''\right].
\label{eq:sigmatt}
\end{equation}
The first term in the above equation is the so-called diamagnetic term, which is proportional to the expectation value of the 1D stress tensor operator $\tau=t_h\sum\limits_{i,\sigma}(c_{i,\sigma}^{\dagger}c_{i+1,\sigma}+\text{H.c.})$. The second term is the two-time susceptibility
\begin{equation}
\chi(t',t'')=-\mathrm{i}\theta(t'-t'')\eval{\psi(t)}{[j^{I}(t'),j^{I}(t'')]}{\psi(t)},
\label{eq:chi}
\end{equation}
where the interaction representation of the current operator reads $j^{I}(t')=U^\dagger(t',t)\,j\,U(t',t)$, with $U(t',t)$ to be the time-evolution operator in the absence of probing perturbations. Note that, due to the absence of time-translation invariance, the formalism is always marked by two time arguments. More details can be found in Ref.~\cite{Zala14}.

Before closing this section, we have a few words for the time-dependent Lanczos method, which is employed to trace the evolution of a wave function under the influence of a Hamiltonian $H(t)$. The key formula is~\cite{Prelovsek}
\begin{equation}
\ket{\psi(t+\delta{t})}=e^{-\mathrm{i}H(t)\delta t}\ket{\psi(t)}
\simeq\sum_{l=1}^{M}{e^{-\mathrm{i}\epsilon_l\delta{t}}}\ket{\phi_l}\braket{\phi_l}{\psi(t)},
\label{eq:lanczos}
\end{equation}
where $\epsilon_l$ and $|\phi_l\rangle$ are the eigenvalues and eigenvectors of the $M$-dimensional Krylov subspace generated in the Lanczos process, respectively. Note that in the Taylor-series expansion of $e^{-iH(t)\delta t}$,  the powers of $H(t)$ that determine the orders of the Lanczos iteration, match those of $\delta t$. It means that with smaller $\delta t$, the required size of the Krylov space $M$ can be reduced. For the time step $\delta{t}=0.02$, we choose $M=30$ to ensure the convergence of numerical results.

We set the on-site repulsion $U=10.0$ in order to restrict our system in the strong-coupling regime. The central frequency of the pumping pulse $\omega_0$ in Eq.~(\ref{eq:vpotent}) is always tuned to match the leading absorption peak in the equilibrium optical conductivity. The results are presented mainly on lattice size $L = 10$ (except for finite-size scaling analysis). In the following discussions, the variable $t$ in both $I(k, \omega, t)$ and $\sigma(\omega,t)$ are usually denoted as $\Delta t$, specifying the difference between the measuring and pumping times.

\section{Results and discussions}\label{sec_pump}
\subsection{$I(k,\omega)$ in equilibrium}

\begin{figure}
\centering
\includegraphics[width=0.5\textwidth]{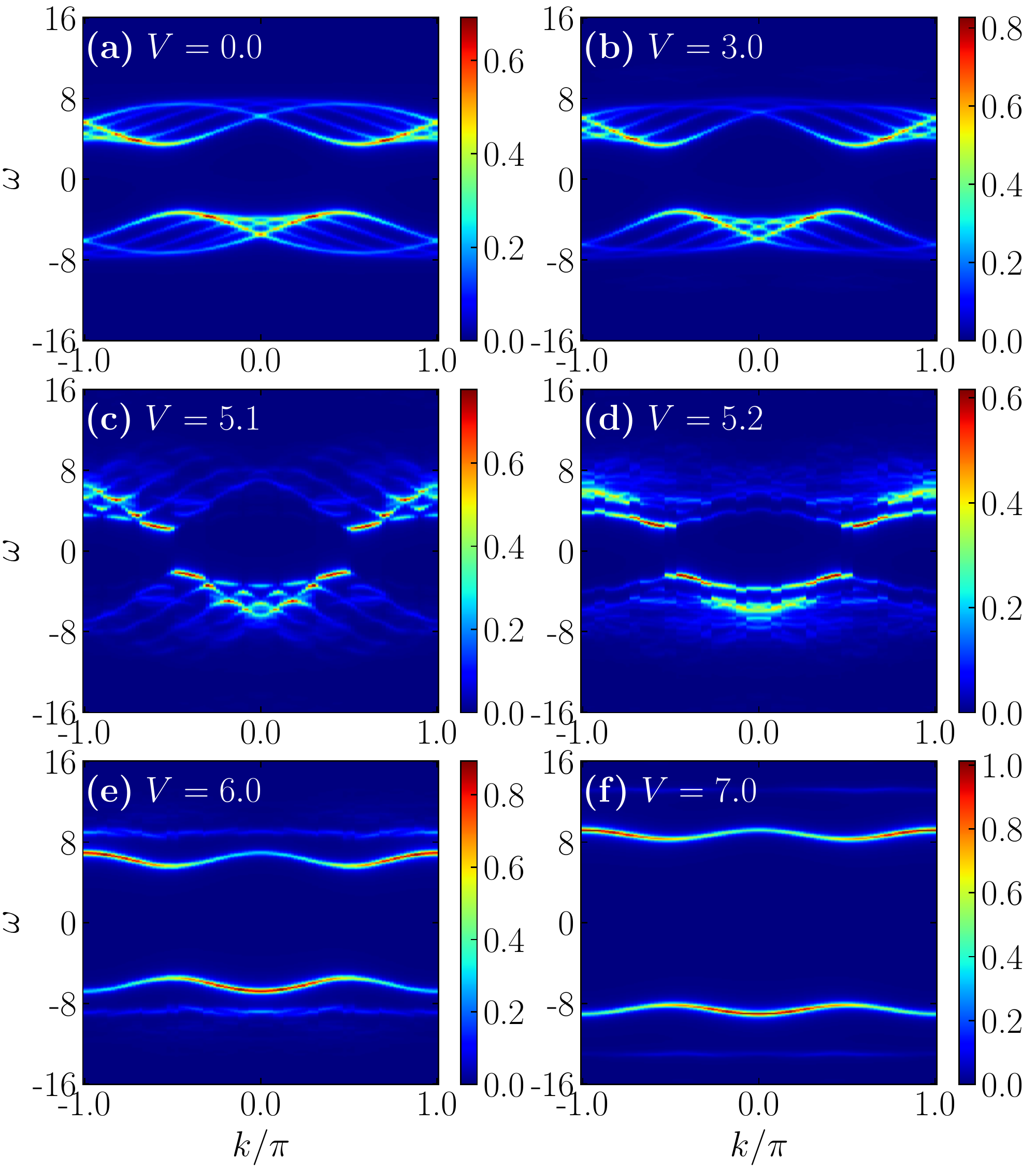}
\caption{(Color online) The equilibrium single-particle spectral function $I(k,\omega)$ for the half filled EHM with $U=10.0$, and (a) $V=0.0$, (b) $V=3.0$, (c) $V=5.1$, (d) $V=5.2$, (e) $V=6.0$, and (f) $V=7.0$, respectively.}
\label{fig_equilibrium}
\end{figure}

\begin{figure}
\centering
\includegraphics[width=0.5\textwidth]{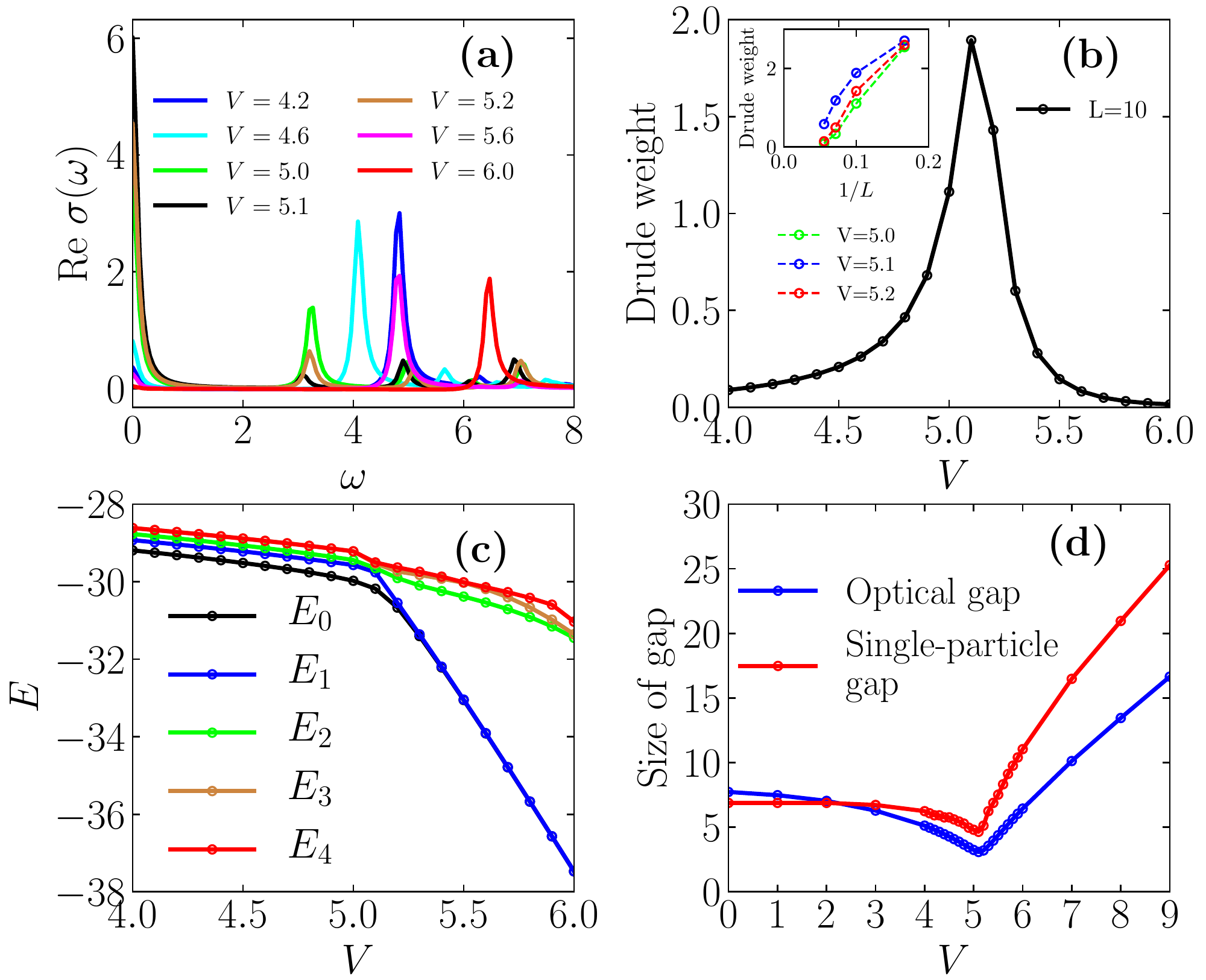}
\caption{(Color online) (a) The real part of the zero-temperature optical conductivity $\RE\,\sigma(\omega)$ of the half filled EHM with $U=10.0$ and $V$ ranging from $4.2$ to $6.0$. (b) The estimation of the Drude weight obtained by the optical sum rule. The inset shows the finite-size scaling for $V=5.0, 5.1$ and $5.2$ with $L=6,10,14$ and $18$. (c) Five lowest-lying energy levels with respect to the change of $V$ from $4.0$ to $6.0$ for $L=10$. (d) The comparison of the $V$ dependence of the optical gap obtained from $\RE\,\sigma(\omega)$ and the single-particle gap estimated by $I(k,\omega)$. To pin down the critical point, the $V$ spacing is set to be $0.1$ for $V\in[4.0,6.0]$.}
\label{fig_equilibrium2}
\end{figure}

Before going to the issue of pump-probe dynamics, let us first examine $I(k,\omega)$, the single-particle spectral function in equilibrium (zero temperature) for the half filled EHM. The results with $L=10$, $U=10.0$ and several $V$'s are shown in Fig.~\ref{fig_equilibrium}, which cover the SDW phase and the CDW phase in the strong-coupling regime. We note that the spectra below (above) the Fermi energy $E_F$, i.e., $\omega=0$, are exclusively composed of $I_{-}(k,\omega)$ [$I_{+}(k,\omega)$], in which a Lorentzian broadening of $0.2$ is introduced in the $\omega$ space. With increasing $V$ from $0.0$ to $5.1$ in Figs.~\ref{fig_equilibrium}(a)-\ref{fig_equilibrium}(c), the single-particle gap decreases. With further increasing $V$ in Figs.~\ref{fig_equilibrium}(d)-\ref{fig_equilibrium}(f), the gap increases. The minimum gap size is observed at $V=5.1$, which indicates a first-order phase transition from SDW to CDW. More details can be found in the later discussions on Fig.~\ref{fig_equilibrium2}.

From Fig.~\ref{fig_equilibrium}, we can also observe that the single-particle spectra show distinct features in the SDW and CDW phases. In SDW with small nearest-neighbor repulsion $V$ as shown in Figs.~\ref{fig_equilibrium}(a) and \ref{fig_equilibrium}(b), the upper and lower Hubbard bands are composed of some interlaced ``stripes". The ``striped" structure is due to finite-size effects and has been addressed in Refs.~\onlinecite{Kim96, Aichhorn04, Kim06}. What happens in the thermodynamic limit is that the ``stripes" will develop into two spinon and holon branches, as the consequence of spin-charge separation in 1D electronic systems~\cite{Claessen02, Sing03, Kim98, Kim06}. The fundamental structure in the SDW side is preserved even close to the phase boundary, as indicated in Fig.~\ref{fig_equilibrium}(c).

The spectral features in CDW are different from those in SDW. For $V=5.2$, the upper (lower) Hubbard band is substantially split into two separated bands as the system moves into the CDW phase [Fig.~\ref{fig_equilibrium}(d)]. For convenience, we focus on the two bands below the Fermi energy $E_F$ and refer to the one closer to the Fermi surface as band I and the other as band II (similar discussions can be applied to the bands above the Fermi energy). With further increase of $V$, the two bands become flatter, accompanied by a redistribution of the spectral weight. More specifically, from Figs.~\ref{fig_equilibrium}(e) and \ref{fig_equilibrium}(f), we can see that the band I gains more weight and becomes brighter, while the band II loses weight and becomes darker. At $V=7.0$ deep into the CDW phase, the band II is almost invisible with only a small amount of spectral weight remaining. We have checked larger systems with size $L=14$ and got consistent results. We will return to the issue of the origin of the two-band structure in a later discussion.

To examine the critical-like features near the SDW-CDW phase boundary, we provide in Fig.~\ref{fig_equilibrium2} some detailed analysis. First, in Fig.~\ref{fig_equilibrium2}(a), the real parts of the optical conductivity $\RE\,\sigma(\omega)$ are presented for $V\in[4.2,6.0]$. The conductivity is calculated by a pump-probe method described in Ref.~\onlinecite{Shao16}, which is capable of capturing a critical behavior at $\omega=0$ (i.e., the Drude component). Here we would like to note that even for an insulator, the Drude weight $D$ can be finite if the size of the system with periodic BC is small~\cite{Fye1991}. For our system of $L=10$, the nonzero Drude component near the phase boundary ($V\in[4.2,6.0]$) can be read in Fig.~\ref{fig_equilibrium2}(a).

The Drude weight can be either estimated from the integration of $\RE\,\sigma(\omega)$ over the zero-frequency peak shown in Fig.~\ref{fig_equilibrium2}(a), or evaluated, more accurately, by the optical sum rule (e.g., see the relevant discussions in Refs.~\onlinecite{Fye1991, Dagotto1994}). Here we apply the sum-rule method for the Drude weight calculation and the results are summarized in Fig.~\ref{fig_equilibrium2}(b). The main figure displays the $V$ dependence of the Drude weight $D$ for the system of $L=10$ and $U=10.0$, with the maximum located at $V=5.1$. While in the thermodynamic limit, as we know the Drude weight should go to zero everywhere, including even the {\em first-order} phase-transition point~\cite{Sandvik04}. The inset in Fig.~\ref{fig_equilibrium2}(b) shows the finite-size scaling of $D$ for three values of $V$ (i.e., $V=5.0$, $5.1$ and $5.2$) in the vicinity of the phase boundary with $L=6,10,14$ and $18$. It is easy to find that as $L$ increases all the resulting $D$'s approach zero.

The transition from SDW to CDW phase can also be unveiled by the change in the low-lying spectra. In Fig.~\ref{fig_equilibrium2}(c), five lowest-lying energy levels as a function of $V\in[4.0,6.0]$ are plotted. We can see that the inflection points on the curves of the ground-state energy $E_0$ and the first-excited energy $E_1$, are both located in the vicinity of $V=5.1$. After $V=5.3$, the two energy levels merge together, indicating a twofold degenerate ground state with the CDW order. Figure~\ref{fig_equilibrium2}(d) shows the $V$ dependence of the single-particle gap [obtained from $I(k,\omega)$] and the optical gap [determined by the position of the first optical absorption peak in $\RE\,\sigma(\omega)$]. The minimums of the two gaps are situated at $V=5.1$. We note that the difference between the two gaps can be attributed naturally to the presence of excitonic effect in the optical conductivity, which is absent in the single-particle spectrum $I(k,\omega)$. From the results in Fig.~\ref{fig_equilibrium2}, we conclude that for the $10$-site system with $U=10.0$, it is most ``critical" when $V=5.1$. This conclusion is consistent with the phase diagram of this model at half filling~\cite{Sandvik04, Ejima07}.


\subsection{$I(k,\omega,t)$ after pump}

\begin{figure}
\includegraphics[width=0.5\textwidth]{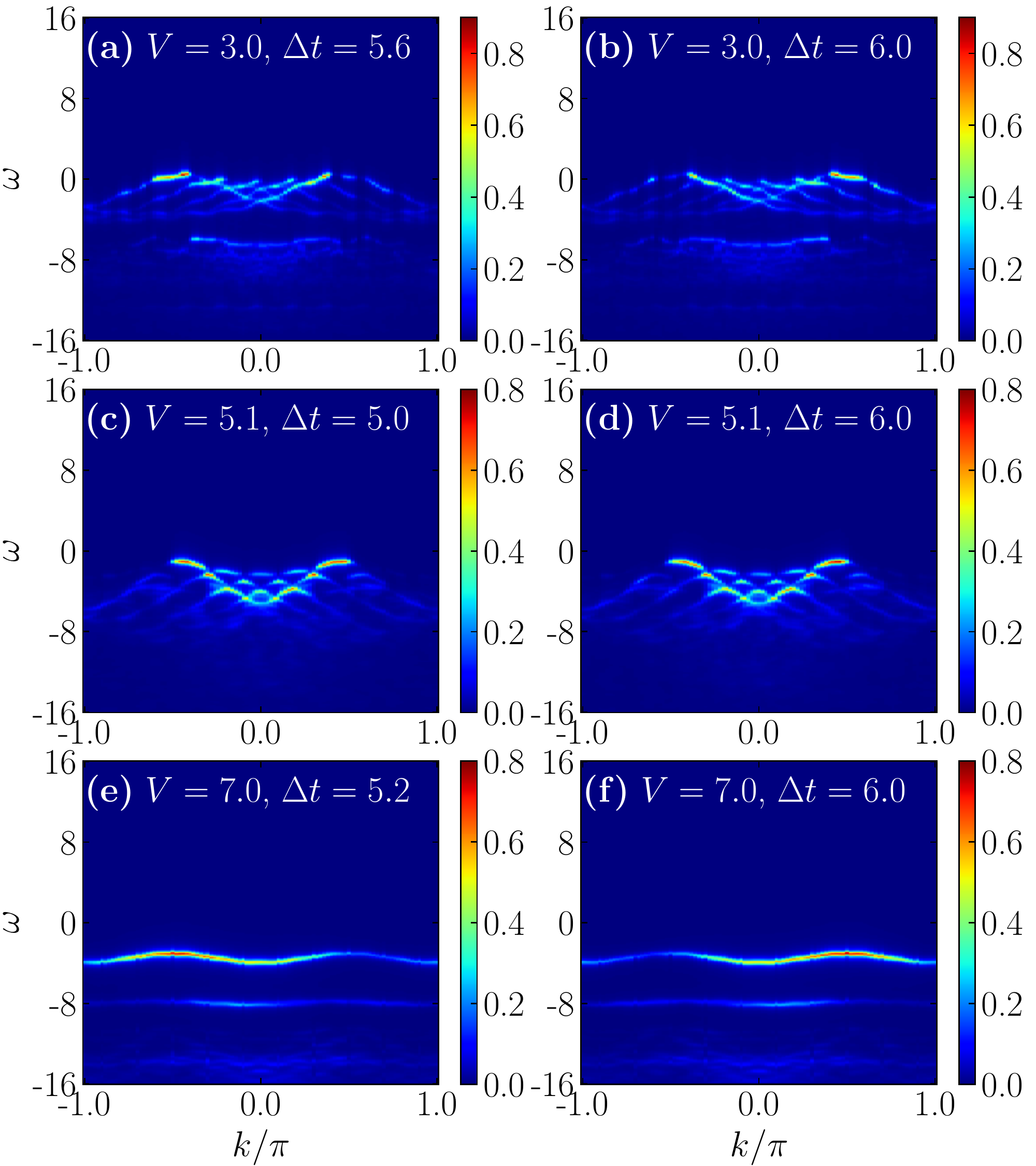}
\caption{(Color online) $I_{-}(k, \omega, \Delta t)$ of the half filled EHM for $U=10.0$, and (a) \& (b) $V=3.0$, (c) \& (d) $V=5.1$, (e) \& (f) $V=7.0$. $\Delta t$ is the time difference between the probing time $t$ and the central time of pumping pulse $t_0$. For each $V$, we select two typical $\Delta t$'s in order to unveil possible spectral weight oscillations. Parameters of the pumping pulse: $A_0 = 0.3$, $t_d = 0.5$.}
\label{fig_pump}
\end{figure}

We are now ready to move to the nonequilibrium case where the system is driven by a transient laser pulse described by Eq.~(\ref{eq:vpotent}). Figure~\ref{fig_pump} shows the time-dependent single-particle spectral function at two different $\Delta t$'s for each $V$, where $\Delta t$ is the difference between the probing time and the pumping time. The pump amplitude $A_0=0.3$, and the pulse width $t_d=0.5$. For the sake of clarity, only the electron-removal part $I_{-}(k, \omega, \Delta t)$ is displayed since the electron-addition part $I_{+}$ is simply a reflection of $I_{-}$ with respect to $\omega=0$ and $k\leftrightarrow \pi-k$. There are several features we would like to bring to the reader's attention.

First of all, we see that, after applying the pumping pulse, the electron-removal (lower Hubbard) bands move upwards as a whole due to the injected energy, since $E_0^{\kappa}$ in Eqs.~(\ref{A+}) and (\ref{A-}) for $I_{+}$ and $I_{-}$ has to be replaced by $E^{\kappa}(t)$ in a nonequilibrium situation. For example, in Figs.~\ref{fig_pump}(a) and \ref{fig_pump}(b) with $V=3.0$ after pump, the bands have already touched the Fermi level with a distorted ``stripe" feature. These are characteristics of the photoinduced insulator-to-metal transition~\cite{Petersen11, Perfetti06}. Second, for both $V=3.0$ and $7.0$, there appear some amount of spectral weights separating from the principal structure and moving downwards along the $\omega$ direction. That is to say, the weight distributions are getting more diverse. In the case of $V=7.0$, the weight transfer mainly takes place between band I and band II. Third, by investigating the temporal profile of $I_{-}(k,\omega,\Delta t)$, we notice that there are regular spectral weight oscillations for $V=3.0$ and $7.0$ with specific periods. They can be readily recognized from the corresponding figures in Fig.~\ref{fig_pump}~\footnote{Note that these $\Delta t$'s we have chosen do not necessarily correspond to the characteristic periods.}. However, there is no similar oscillation that can be identified for $V=5.1$ [Figs.~\ref{fig_pump}(c) and \ref{fig_pump}(d)]. The reason will be discussed later.

\subsubsection{Leading oscillation and optical gap}

\begin{figure}
\includegraphics[width=0.5\textwidth]{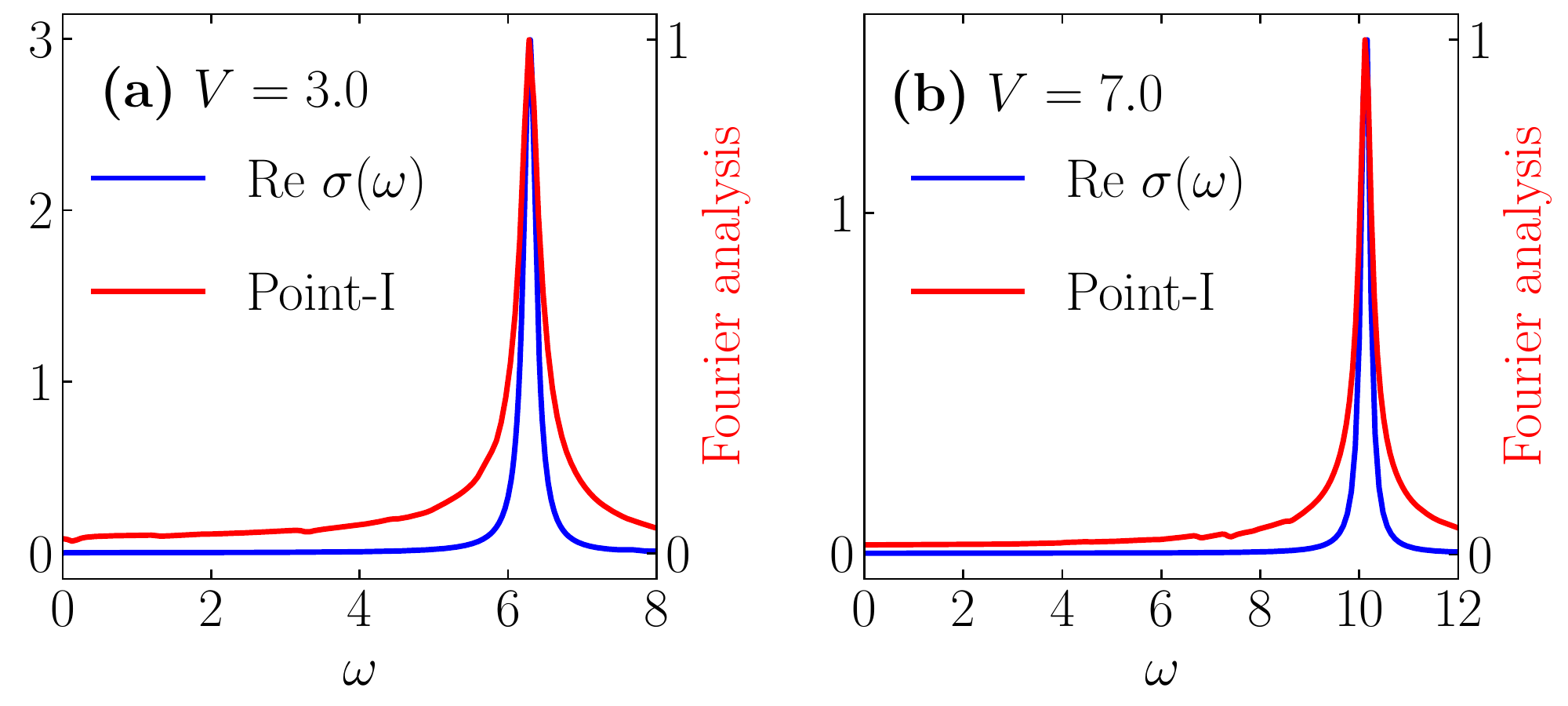}
\caption{(Color online) Comparison between the equilibrium optical conductivity $\RE\,\sigma(\omega)$ and the Fourier transformation of the magnitude of $I_{-}(k,\omega,\Delta t)$ at point I for the half filled EHM for $U=10.0$, and (a) $V=3.0$ and (b) $V=7.0$ (see text for detail). Parameters of the pumping pulse are $A_0 = 0.3$, $t_d = 0.5$.}
\label{fig_pointI}
\end{figure}

To study the oscillations shown in Fig.~\ref{fig_pump}, we need to fix some typical points on the bands and perform the Fourier transformation of their spectral intensities. We first focus on the two points on the lower Hubbard band that are closest to the Fermi level. They are located at $k=\pm\pi/2$ (with corresponding $\omega$ values) and turn out to be the most weighted points. We can call either of them point I (note that for the case of $V=7.0$, point I is located on band I). More specifically, we chose $I_{-}(k=\pi/2,\,\omega=0.16,\,\Delta t)$ for $V=3.0$ and $I_{-}(k=\pi/2,\,\omega=-3.12,\,\Delta t)$ for $V=7.0$ to perform the Fourier transformations with $\Delta t\in[5,105]$.  The results are presented in Fig.~\ref{fig_pointI} in red lines, with the equilibrium optical conductivity displayed in blue lines. We find that the peak of the Fourier amplitude matches the position of the main optical absorption peak $\omega_c$ quite well.

Since the optical conductivity measures the charge transport ability under the stimulus of an AC electric field, it is quite natural to expect that the leading oscillation frequency of the single-particle spectrum should match the intrinsic charge gap $\omega_c$. The argument can be reinforced by the observation that, even if the pumping frequency $\omega_0$ [in Eq.~(\ref{eq:vpotent})] {\em deviates} from $\omega_c$, the frequency of the induced oscillation of the single-particle spectrum does not move away from $\omega_c$ (not shown here).

Now we are in a position to explain the absence of the spectral oscillation in $V=5.1$. This is due to the presence of the significant zero-frequency peak (the Drude weight) in the equilibrium optical conductivity $\RE\,\sigma(\omega)$ as shown in Figs.~\ref{fig_equilibrium2}(a) and \ref{fig_equilibrium2}(b). The correspondence between the leading oscillation in the single-particle spectrum and the position of the leading absorption peak in the optical conductivity suggests that, in this case, the possible temporal oscillations in $I(k,\omega,t)$ can be largely wiped off.

\subsubsection{Band-II oscillation and the photoinduced state}

\begin{figure}
\includegraphics[width=0.5\textwidth]{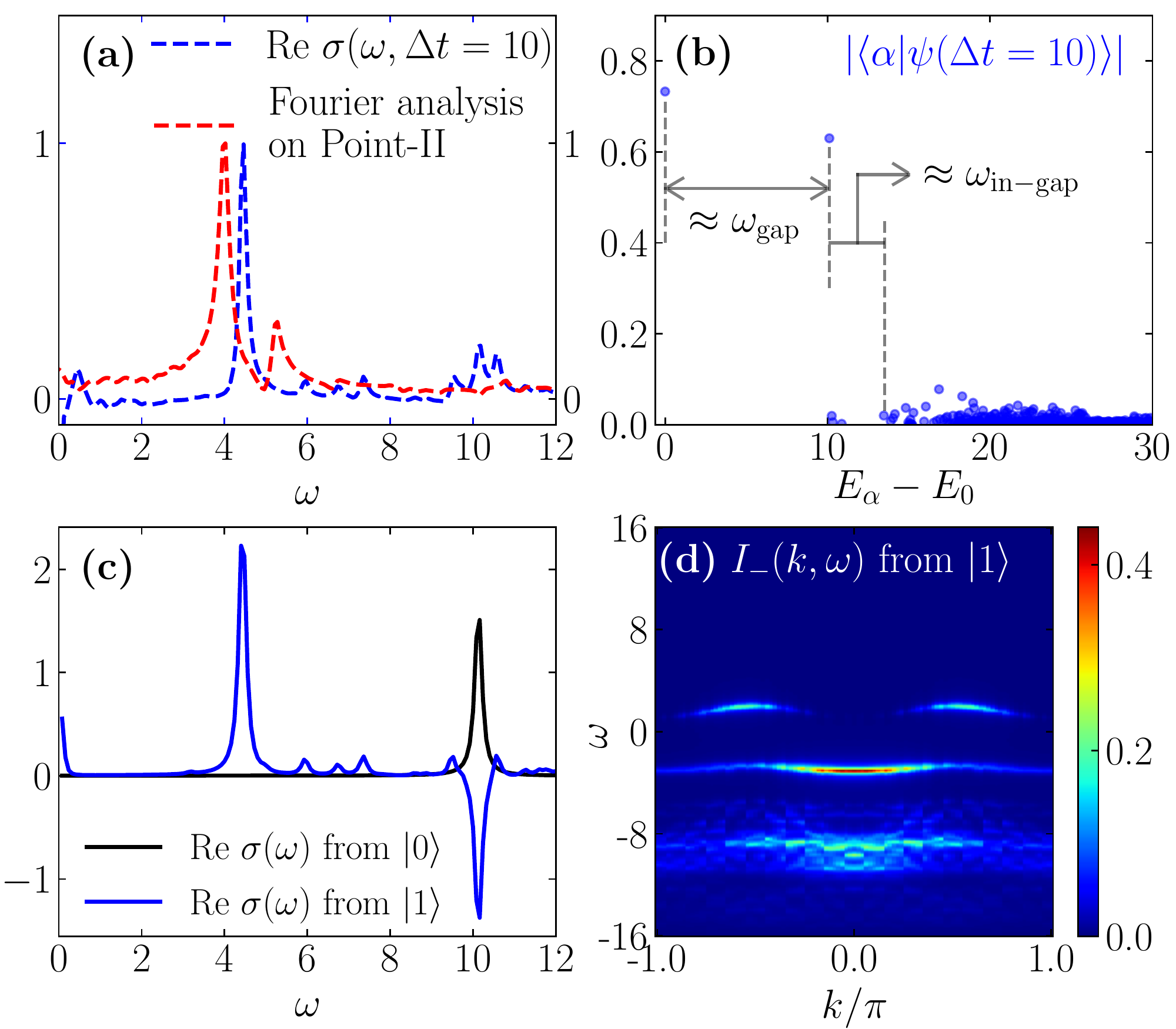}
\caption{(Color online) (a) The comparison between $\RE\,\sigma(\omega, \Delta t)$ and the Fourier transformation of $I_{-}(k,\omega,\Delta t)$ with respect to $\Delta t\in[5,105]$ at point II. (b) Overlap between $\ket{\Psi(\Delta t=10)}$ and the eigenstates $\ket{\alpha}$ as the function of $E_{\alpha}-E_0\,\in[0,30]$. (c) $\RE\,\sigma(\omega)$ calculated with respect to the ground state $\ket{0}$ (black line) and the first-excited state $\ket{1}$ (blue line), respectively. (d) $I_{-}(k,\omega)$ calculated with respect to the first-excited state $\ket{1}$. Parameters of the EHM: $U=10.0$, $V=7.0$. Parameters of the pumping pulse: $A_0 = 0.3$, $t_d = 0.5$.}
\label{fig_pointII}
\end{figure}

In this section we focus on the two-band structure of the single-particle spectral function for $V=7.0$ in the CDW phase. Recall that the signal of band II is enhanced by the optical pump, and the oscillation of its spectral weight can be perceived [see Figs.~\ref{fig_pump}(e) and \ref{fig_pump}(f)]. To quantify it, we choose the point on the band II curve with $k=0$ and the corresponding energy $\omega=-8.16$ (denoted point II) for detailed investigations. The intensity of point II is found to be enhanced from $0.05$ to $0.19$ (average) by the given pump. The Fourier transform of its oscillation is shown in Fig.~\ref{fig_pointII}(a) with dashed red curve, where the main peak at $\omega=4$ is identified. Interestingly, very close to the frequency of $\omega=4$, a pump-induced peak in $\RE\,\sigma(\omega,\Delta t = 10)$ can be found as shown by the dashed blue curve in Fig.~\ref{fig_pointII}(a). This peak appears inside the optical gap [as read from Re $\sigma(\omega)$ from $\ket{0}$ in Fig.~\ref{fig_pointII}(c)] and indicates a photoinduced in-gap excitation~\cite{Lu15}. What is the relation between the enhancement of band II and the emerging photoinduced excitation in $\RE\,\sigma(\omega,\Delta t)$? We will address this issue in the remaining discussion of this section.


First, the nature of the in-gap excitation can be easily understood from Fig.~\ref{fig_pointII}(b), where the overlaps between the evolved wave function at the given time $\Delta t =10$, and all the eigenstates $\ket{\alpha}$'s with $E_{\alpha}-E_0\in[0,30]$ are plotted. (Note that, after turning off the pumping pulse, the magnitudes of these overlaps remain unchanged.) We find that $\ket{\Psi(\Delta t=10)}$ has very large overlap with the ground state $\ket{0}$ and the first-excited state $\ket{1}$. In other words, $\ket{\Psi(\Delta t=10)}$ is essentially described by a superposition of $\ket{0}$ and $\ket{1}$. Based on this fact, we calculate the optical conductivity directly form $\ket{1}$ as shown by blue curve in Fig.~\ref{fig_pointII}(c), which has a peak at $\omega\approx4.4$, being the same energy for the photoinduced in-gap peak in Fig.~\ref{fig_pointII}(a). This means that the latter can be attributed to an optical excitation from $\ket{1}$. As a consequence, the positive peak with $\omega\approx4.4$ corresponds to the optical absorption from $\ket{1}$ to another optically allowed state with higher energy, as indicated by the symbol ``$\omega_{\mathrm{in-gap}}$'' in Fig.~\ref{fig_pointII}(b).

In Fig.~\ref{fig_pointII}(c), a peak with negative weight is located at $\omega\approx10.1$, which comes from an optical emission process from $\ket{1}$ back to $\ket{0}$ via dipole transition. Since the energy of this peak is the same as that of the absorption peak from $\ket{0}$ at $\omega=10.12$, the suppression of spectral weight at $\omega\approx10.1$ in $\RE\,\sigma(\omega, \Delta t=10)$ is naturally understood as a consequence of the two opposite contributions, as shown in Fig.~\ref{fig_pointII}(a).

With the experience in the time-resolved optical conductivity, we imagine that in the time-resolved single-particle spectrum $I(k,\omega,t)$, there should also be the contribution from $\ket{1}$. This point is supported by the calculation of the electron-removal part $I_{-}(k,\omega)$ directly from $\ket{1}$, as shown in Fig.~\ref{fig_pointII}(d). By comparing with Figs.~\ref{fig_pump}(e) and \ref{fig_pump}(f), we deduce that the enhanced band II there (around $\omega\sim -8.0$) mainly comes from the bright and narrow band in Fig.~\ref{fig_pointII}(d) (in the vicinity of $\omega\sim -3.0$). We note that the mismatch of the $\omega$ values here is due to different energy displacements in calculating $I_{-}(k,\omega,\Delta t)$ and $I_{-}(k,\omega)$ from $\ket{1}$: the difference between the energy of $\ket{1}$ and the energy of an injected pump photon is just $5$. Further discussion on the relation between band II and the photoinduced in-gap excitation from the perspective of the pump strength $A_0$ dependence can be found in the appendix.


Let us make a brief summary of results regarding band II up to now. First recall that, when the phase boundary is crossed from the SDW to the CDW side, the single-particle spectrum evolves from the stripe structure into a well-identified two-band structure in the vicinity of the phase boundary [Figs.~\ref{fig_equilibrium}(c) and \ref{fig_equilibrium}(d)]. With $V$ increasing further, band II gradually diminishes. It is already quite weak when $V=7.0$ [Figs.~\ref{fig_equilibrium}(e) and \ref{fig_equilibrium}(f)]. However, accompanied by the photoinduced excitation of the first-excited state, the signal of band II is reinvigorated [Figs.~\ref{fig_pump}(e) and \ref{fig_pump}(f)], and its spectral oscillation coincides with the optical signal of the excited state (Fig.~\ref{fig_pointII}(a)). The properties of the first-excited state have been addressed with a long-range BOW order identified~\cite{Shao19}. Based on the above evidence, we propose that band II can be understood as the single-particle spectrum in the bond-order background. More interestingly, it can be enhanced with its own characteristic oscillation by a proper pumping pulse.

As a result, we propose that, instead of measuring the time-resolved optical conductivity (or reflectivity) in the low-frequency regime, which can be out of reach in the present techniques, the trARPES experiment on relevant materials can be employed to detect the existence of the hidden BOW order in the excited states via the time-resolved single-particle spectrum.

\subsection{Time-dependent momentum distribution function}\label{subsec:MDF}

\begin{figure}
\centering
\includegraphics[width=0.5\textwidth]{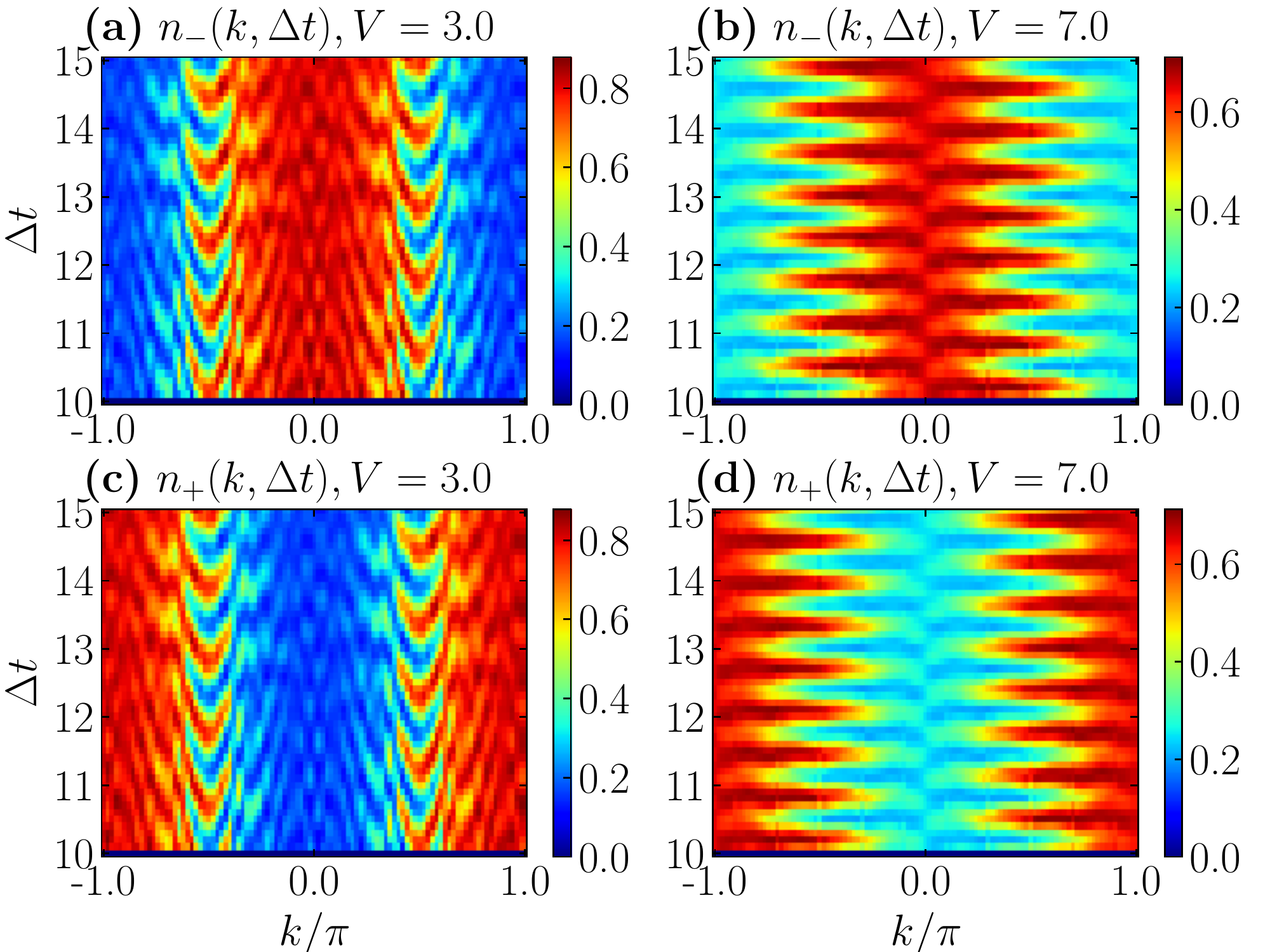}
\caption{(Color online) Contour plot of time-dependent momentum distribution functions $n_{-}(k,\Delta t)$ [$n_{+}(k,\Delta t)$] for $V=3.0$ (a) [(c)] and $V=7.0$ (b) [(d)] with $U=10.0$, respectively. Parameters of the pumping pulse are $A_0 = 0.3$, $t_d = 0.5$.}
\label{fig_nk}
\end{figure}

Before closing the section, we present the results on the time-dependent momentum distribution functions (TDMDs) of holes and electrons. The TDMD has its experimental significance, since it can be measured via the trARPES data~\cite{Randeria1995} or time-resolved Compton scattering~\cite{Kemper2013} in condensed-matter experiments, as well as time-of-flight absorption images in cold atom systems. In theory, they are defined as
\begin{eqnarray}
n_{+}(k,\Delta t)=\int_{-\infty}^{+\infty} I_{+}(k,\omega,\Delta t)\,d \omega
\label{n+}
\end{eqnarray}
and
\begin{eqnarray}
n_{-}(k,\Delta t)=\int_{-\infty}^{+\infty} I_{-}(k,\omega,\Delta t)\,d \omega.
\label{n-}
\end{eqnarray}
Note that, due to the sum rule of the spectral function, $n(k,\Delta t):=n_{+}(k,\Delta t)+n_{-}(k,\Delta t)$ should be equal to $1$ for any values of $k$ and $\Delta t$. The results of $n_{+}(k,\Delta t)$ and $n_{-}(k,\Delta t)$ for $V=3.0$ and $V=7.0$ are given in Fig.~\ref{fig_nk}.

We find that the oscillation of the momentum distribution in SDW is mainly confined in the vicinity of $k=\pm\pi/2$, while it is more dispersive in the CDW phase. These features are consistent with the behavior of $I_{-}(k,\omega,\Delta t)$ shown in Fig.~\ref{fig_pump}, where one can notice that the oscillation on band I in CDW [Figs.~\ref{fig_pump}(e) and \ref{fig_pump}(f)] has a wider momentum distribution than the case of SDW. Additionally, in Figs.~\ref{fig_nk}(b) and \ref{fig_nk}(d), there are oscillations taking place in the vicinity of $k=0$, which come from the contribution of band II and have different frequencies compared with oscillations around $k=\pm\pi/2$. We have checked that the Fourier component of $n_{-}(k=0,\Delta t)$ in CDW has a main peak around $\omega=4$ which is consistent with the direct Fourier transform of the oscillation on band II [Fig.~\ref{fig_pointII}(a)]. It means that the characteristic oscillations in the single-particle spectrum addressed in the above sections can also be resolved even without energy resolution. We then conclude that the TDMD can provide an alternative way to identify the photoinduced BOW state.


\section{Conclusion}\label{sec_conclusion}

In this paper, by combining twisted BCs with the time-evolution Lanczos technique in ED calculations, we have presented a study of the time-resolved single-particle spectral function of the 1D extended Hubbard model at half filling when it is driven by a transient laser pulse. We found that the frequencies of the characteristic oscillations of the spectral weight coincide with the resonant positions of the optical conductivity, which is expressed by the two-particle correlations and reflects the charge transport ability. We have further investigated the evolution of the two-band structure in the CDW phase under photo-irradiation, and argued that band II can be regarded as the spectrum of the single-particle excitation arising from the bond-order background. This can deepen our current understanding from the dynamic perspective of the intermediate BOW state caused by the competition of many-body effects.

The issue remains open about whether this characteristic oscillation in the time-resolved single-particle spectrum can be accessed in trARPES experiments for some materials. In our calculations, the units of energy and time are set by the hopping constant $t_h$. Suppose $t_h=0.1\,\mathrm{eV}$, the time unit turns out to be $6.58\,\mathrm{fs}$. The period for the band II oscillation in our parameter settings is around $10\,\mathrm{fs}$. In view of the current trARPES resolutions (for example, the energy and time resolutions are sub $150\,\mathrm{meV}$ and $30\,\mathrm{fs}$ in Ref.~\onlinecite{EICH2014231}; in Ref.~\onlinecite{Rohde2016} they are $170\,\mathrm{meV}$ and $13\,\mathrm{fs}$), this task is difficult but not impossible with proper materials available. Furthermore, if one sacrifices the energy information while maintaining high momentum resolution, as discussed in Sec.~\ref{subsec:MDF}, the time resolution can be further enhanced and we may have a better chance to resolve the oscillations. Another possibility is to simulate the system directly in cold-atom optical lattices, where one can measure the TDMD with much longer time-scales.


\begin{acknowledgments}
C.S. and H.L. acknowledge discussions with R. Mondaini, S. Tarat, and T. Cadez. We would like to thank the anonymous referees for suggesting the calculation that led to Fig.~\ref{fig_equilibrium2} and the discussion of the $A_0$ dependence in the appendix.
C.S. acknowledges support from the China Postdoctoral Science Foundation (Grant No. 2019M650464) and the NSAF (Grant No. U1530401). H.L. and H.-G.L. acknowledge support from the National Natural Science Foundation of China (NSFC; Grants No. 11474136, No. 11674139, and No. 11834005) and the Fundamental Research Funds for the Central Universities.
T.T. is partly supported by MEXT, Japan, as a social and scientific priority issue (creation of new functional devices and high-performance materials to support next-generation industries; CDMSI) and exploratory challenge (challenge of basic science - exploring extremes through multiphysics and multiscale simulations)  on a post-K computer and by CREST (Grant No. JPMJCR1661).
\end{acknowledgments}


\appendix*
\section{Dependence on Pump Parameter $A_0$}\label{appendix}

\begin{figure}
\centering
\includegraphics[width=0.5\textwidth]{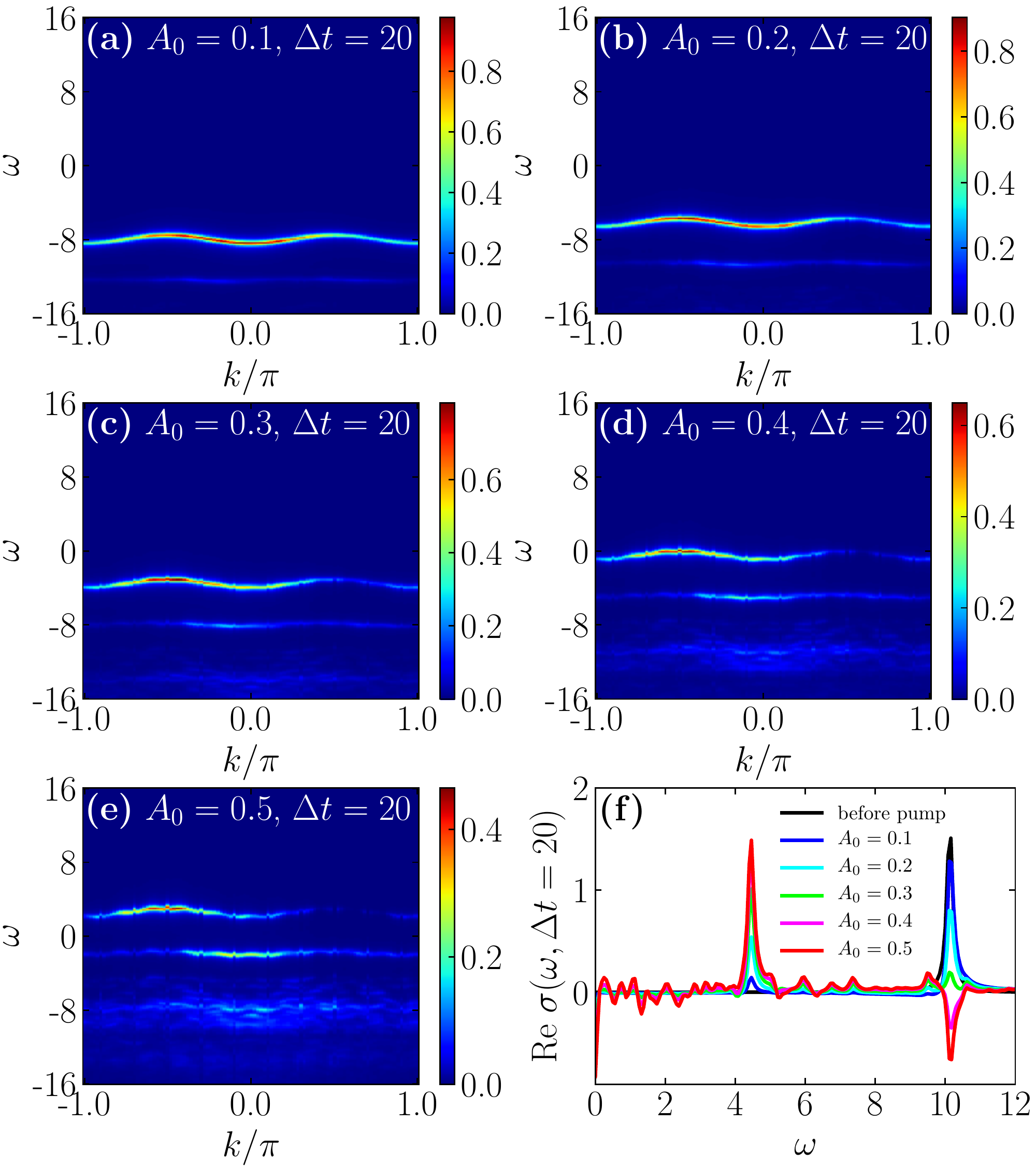}
\caption{(Color online) $I_{-}(k, \omega, \Delta t=20.0)$ of the half filled EHM for $U=10.0$, $V=7.0$ with $A_0=0.1-0.5$ in panels (a)-(e), respectively. (f) $\RE\,\sigma(\omega)$ before and after pump at $\Delta t=20$ with different $A_0$. Parameters of the pumping pulse: $\omega_0=10.12$, $t_d = 0.5$.}
\label{fig_A1}
\end{figure}

\begin{figure}
\centering
\includegraphics[width=0.5\textwidth]{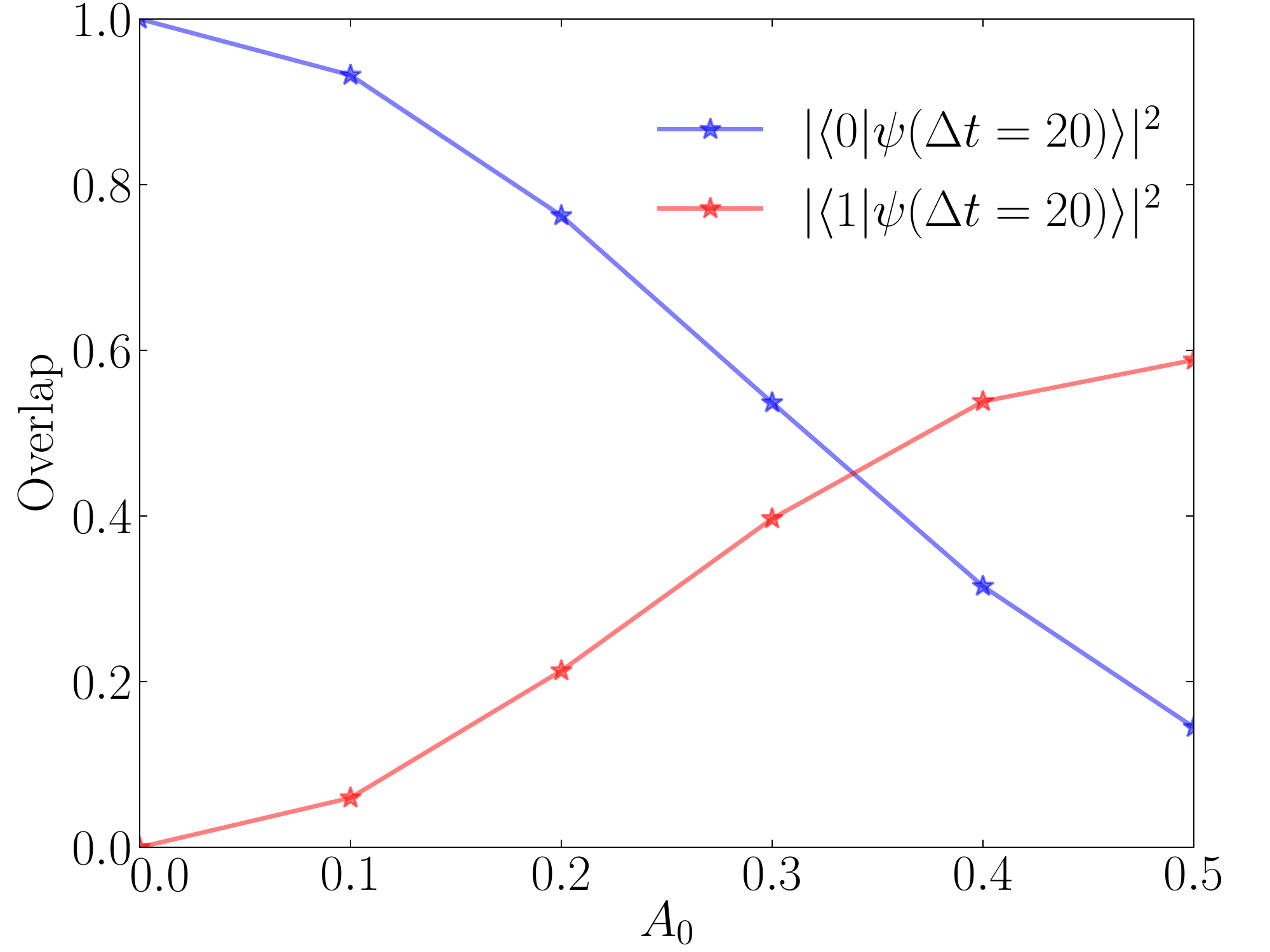}
\caption{(Color online) The square of the overlap between the ground-state (the first-excited state) and the time-evolving wave function at $\Delta t=20$ as a function of the strength of pumping pulse $A_0$, shown in blue curve (red curve).}
\label{fig_A2}
\end{figure}

In the main text, we have chosen the strength of pumping pulse $A_0=0.3$ for the demonstration. In this appendix, we present the influence of $A_0$ for the half filled EHM with $U=10.0$ and $V=7.0$ in detail, and the results are summarized in Figs.~\ref{fig_A1} and \ref{fig_A2}.

In Figs.~\ref{fig_A1}(a)-\ref{fig_A1}(e), the time-resolved electron-removal spectral function $I_{-}(k,\omega,\Delta t=20.0)$ under various $A_0$ is shown. We can see that, upon increasing $A_0$ from $0.1$ to $0.5$, the signature of band II is enhanced. We note that the upward shift of the whole electron-removal bands is due to the optically injected energy: in the calculation of $I_{\pm}(k,\omega,t)$, $E_0^{\kappa}$ in Eqs.~(\ref{A+}) and (\ref{A-}) is replaced by $E^{\kappa}(t)=\eval{\Psi^{\kappa}(t)}{H}{\Psi^{\kappa}(t)}$, as we mentioned in Sec.~\ref{sec_model}.

Accompanied by the enhancement of band II in $I_{-}(k,\omega,\Delta t)$ is the growing contribution of the first-excited state to the time-evolving wave function $\ket{\psi(t)}$, which can be easily recognized both in the time-resolved optical conductivity and in the overlap calculation, as shown in Figs.~\ref{fig_A1}(f) and \ref{fig_A2}, respectively. The connection between band II, the in-gap excitation in $\RE\,\sigma(\omega,\Delta t)$, and the photoinduced first-excited state can thus be identified.
%

\end{document}